\documentstyle[12pt]{article}
\input amssym.def                    

\newcommand{\identity}[0]{{\Bbb I}}

\begin{document}\addtolength{\baselineskip}{1mm}
\begin{titlepage}
\title{Fokker-Planck-Boltzmann equation for dissipative particle dynamics}
\author{C.A. Marsh\\
Theoretical Physics, Oxford University, 1 Keble Road,\\ 
Oxford OX1 3NP, UK\\ 
and\\ 
G. Backx and M.H. Ernst\\
Institute for Theoretical Physics, Universiteit Utrecht,\\
3508 TA Utrecht, NL}

\label{lns}
\maketitle
\begin{center}
\section*{Abstract}
\end{center}

The algorithm for Dissipative Particle Dynamics (DPD), as modified by Espa\~{n}ol 
and Warren, is used as a starting point for proving an $H$-theorem for the free 
energy and deriving hydrodynamic equations.
Equilibrium and transport properties of the DPD fluid are explicitly calculated in 
terms of the system parameters for the continuous time version of the model. 

\vspace{0.5cm}

PACS. 05.20 Dd Kinetic theory;
      05.70 Ln Nonequilibrium thermodynamics, irreversible processes;
      47.11+j Computational methods in fluid dynamics.
	
\end{titlepage}
\newpage

In recent years several new simulation methods have been proposed for 
studying dynamical and rheological properties of complex fluids, on time
scales which are difficult to reach by conventional molecular dynamics methods.
These new techniques include lattice gas cellular automata (LGCA), the
lattice Boltzmann equation (LBE), and dissipative particle dynamics (DPD).

The goal of this letter is a theoretical analysis of the equilibrium and
transport properties of a DPD system. This method was introduced
by Hoogerbrugge and Koelman \cite{H+K} and modified by Espa\~{n}ol and
Warren \cite{E+W} to ensure a proper thermal equilibrium state. There
exist many recent applications of this new technique to concentrated 
colloidal suspensions, dilute polymer solutions and phase separation in binary 
mixtures \cite{Colloid}.  Of course, to study the rheology of
colloidal particles and polymers, suspended in a DPD fluid, one needs
to model in addition the hydrodynamic interactions between these
objects, transmitted through the fluid \cite{Colloid}, which is not the 
goal of this article.

We briefly recall the basic elements of the simulation method.
The DPD particle-based algorithm considers N point particles 
which model a fluid out of equilibrium and 
conserve mass and momentum, but not energy. Positions and velocities
are continuous, but time is discrete and incremented
in time steps $\delta t$, as in LGCA and LBE. The algorithm consists of a 
collision step and a 
propagation step. In the {\em collision step}, the velocity of each
particle is updated according to its interaction with
particles inside a sphere of radius $R_0$ through conservative,
frictional and random forces.
In the subsequent {\em propagation} step of fixed length $\delta t$ all
particles move freely to their new positions.

The equations of motion for DPD with a finite step $\delta t$ are given
by \cite{H+K}
\begin{eqnarray} \label{a1}
\mbox{\boldmath $v$}_i(t+\delta t) - \mbox{\boldmath $v$}_i (t) & =&
{\bf a}_i (t,\delta t) \equiv {\sum_{j(\neq i)}} 
\left\{ \mbox{\boldmath $k$}_{ij} \delta t + \mbox{\boldmath $\sigma$}_{ij}
\delta W_{ij} \right\} \nonumber \\
\mbox{\boldmath $r$}_i (t+\delta t) - \mbox{\boldmath $r$}_i (t) & = &
\mbox{\boldmath $v$}_i (t+\delta t)\delta t ,
\end{eqnarray}
where $x_i = \{ \mbox{\boldmath $v$}_i ,\mbox{\boldmath $r$}_i \}$ is the 
phase
of the $i$-th particle $(i = 1,2,\ldots ,N)$. The interparticle force 
contains a systematic part \mbox{\boldmath $k$} and a random part
$\mbox{\boldmath $\sigma$}\delta W$. The systematic force
$\mbox{\boldmath $k$}_{ij} = \mbox{\boldmath $F$}_{ij}/m - 
\mbox{\boldmath $\gamma$}_{ij}$ has a conservative part $\sim
\mbox{\boldmath $F$}$ and a dissipative part $\mbox{\boldmath $\gamma$}$.
The random force is described by a stochastic variable $\delta W$,
defined through
$\delta W_{ij} = \int^{t+\delta t}_{t} d\tau \xi_{ij} (\tau ) ,$
where $\xi_{ij}(t)$ is Gaussian white noise with an average 
$\overline{\xi}_{ij} = 0$ and a correlation
$\overline{\xi_{ij}(t) \xi_{k\ell}(t^\prime )} = ( \delta_{ik} \delta_{j\ell} +
\delta_{i\ell} \delta_{jk}) \delta (t-t^\prime ) ,$
and $\delta (t)$ is a Dirac delta function. The forces are defined as
$\mbox{\boldmath $F$}_{ij} = - \partial\phi (R_{ij} )/\partial
\mbox{\boldmath $r$}_i, 
\mbox{\boldmath $\gamma$}_{ij} = \gamma w_D (R_{ij}) \mbox{\boldmath 
$\widehat{R}$}_{ij}
\mbox{\boldmath $\widehat{R}$}_{ij} \cdot (\mbox{\boldmath $v$}_i -\mbox{\boldmath 
$v$}_j)$ and 
$\mbox{\boldmath $\sigma$}_{ij} = \sigma w_R (R_{ij}) \mbox{\boldmath 
$\widehat{R}$}_{ij},$
where $\phi (R)$ is a pair potential, $w_D(R)$ and $w_R (R)$ are
positive weight functions, and $\gamma$ and $\sigma$ are respectively the
strength of the friction and the noise. The weight function vanishes 
outside a finite range $R_0$ (here chosen to be equal for $w_D$ and $w_R$).
Moreover, $\mbox{\boldmath $R$}_{ij} = \mbox{\boldmath $r$}_i -
\mbox{\boldmath $r$}_j$, and a hat denotes a unit vector. The interparticle
forces $\mbox{\boldmath $k$}_{ij}$, $\mbox{\boldmath $F$}_{ij}$,
$\mbox{\boldmath $\gamma$}_{ij}$ and $\mbox{\boldmath $\sigma$}_{ij}$ are
antisymmetric under interchange of $i$ and $j$, which guarantees the
conservation of total momentum, which is a prerequisite for the
existence of a slowly chainging local fluid velocity and the validity
of the Navier Stokes equation.

Formally, the DPD algoirthm defines a microscopic N particle system.
However, the introduction of noise and dissipation represents a coarse
grained description.  Consequently, the ``DPD particles'' should not
be interpreted as molecules, but as some mesoscopic degree of freedom
of the fluid, referred to as ``lumps.''
If $t_0 \propto 1/\gamma n R^{d}_{0}$ denotes
the characteristic molecular time scale in DPD with $n = N/V$ the number 
density, $d$ the number of dimensions and $\gamma$ the friction constant,
then $t_0$ is considered to be large compared to any molecular time scale.
In this description the {\em dominant} interactions are friction $(\sim \gamma 
)$
and random noise $(\sim \sigma )$, whereas the conservative forces can be
interpreted as {\em weak} forces of relatively long range, which can be
treated perturbatively. At sufficiently small temperature, they are responsible for phase transitions from liquid to crystalline order. Here they will be set 
equal to zero ($\gamma$ large).
The random forces act effectively as repulsive
forces to prevent collapse of DPD particles. 

Define the single particle and pair distribution functions as
\begin{equation} \label{a5}
f(x,t) = \langle \sum_i \delta (x-x_i (t))\rangle ;
\qquad f^{(2)}(x,x^\prime ,t) = \langle \sum_{i\neq j} \delta (x-x_i(t)) 
\delta (x^\prime - x_j(t))\rangle ,
\end{equation}
where $<\cdots>$ represents an average over an initial ensemble of the 
$N$-particle system.
Combination of (\ref{a5}) and (\ref{a1}) then yields:
\begin{eqnarray} \label{a6}
f( \mbox{\boldmath $v$},\mbox{\boldmath $r$}+\mbox{\boldmath $v$} \delta t,
t + \delta t) &=& \langle \sum_i \delta (\mbox{\boldmath $r$}-\mbox{\boldmath 
$r$}_i
(t)) \delta (\mbox{\boldmath $v$}-\mbox{\boldmath $v$}_i (t) -{\bf a}_i (t)) 
\rangle
\nonumber \\
& = & \langle \sum_i [1-\mbox{\boldmath $a$}_i \cdot \mbox{\boldmath $\partial$} + 
\frac{1}{2}
\mbox{\boldmath $a$}_i\mbox{\boldmath $a$}_i \; : \; \mbox{\boldmath 
$\partial$}
\mbox{\boldmath $\partial$} + \ldots ] \delta (x-x_i (t)) \rangle
\end{eqnarray}
where $\mbox{\boldmath $\partial$} = \partial / \partial \mbox{\boldmath $v$}$ is 
a gradient in the $\mbox{\boldmath $v$}$-variable and (:) denotes a double 
contraction.
As the total force $\mbox{\boldmath $a$}_i$ is small for small $\delta t$, the 
argument of the $\delta$-function may be expanded in powers of 
$\mbox{\boldmath $a$}_i$. A subsequent average over the random noise gives, 
\begin{equation} \label{a7}
f(\mbox{\boldmath $v$},\mbox{\boldmath $r$}+\mbox{\boldmath $v$} \delta t,t+\delta 
t) - f(\mbox{\boldmath $v$},\mbox{\boldmath $r$},t) = \delta t {\cal J}_{1}(f) + 
(\delta t)^2 {\cal J}_{2} (f) 
+ {\cal O}((\delta t)^3) + \cdots,
\end{equation}
where the {\em collision} terms ${\cal J}_1 (f)$ and ${\cal J}_2 (f)$ can
be calculated straightforwardly. Here we only quote the dominant term of
${\cal O}(\delta t)$ explicitly:
\begin{equation}\label{a8}
{\cal J}_1(f) = \mbox{\boldmath $\partial$} \cdot \int dx^\prime \mbox{\boldmath 
$k$} (x,x^\prime)
f^{(2)}(x,x^\prime) + \frac{1}{2} \mbox{\boldmath $\partial$}
\mbox{\boldmath $\partial$} \; : \; \int dx^\prime \mbox{\boldmath $\sigma$}
(x,x^\prime ) \mbox{\boldmath $\sigma$}(x,x^\prime) f^{(2)}(x,x^\prime ).
\end{equation}
The term ${\cal J}_2(f)$ contains ${\cal O}(\mbox{\boldmath $\partial$}^n)$ with
$n = 2,3,4$ and involves the higher order distribution functions
$f^{(2)}$ and $f^{(3)}$. Equation (\ref{a7}) constitutes the first
equation of the BBGKY-hierarchy for the DPD fluid with discrete time steps,
relating the change in $f$ to higher order distribution functions.

To obtain a closed kinetic equation for $f(x,t)$, the so-called the 
Fokker-Planck-Boltzmann (FPB) equation, we
assume {\em molecular chaos},
i.e. $f^{(2)}(x,x^\prime ,t) = f(x,t) f(x^\prime ,t)$,
etc. The dominant collision term
(\ref{a8}) contains the systematic frictional force as well as the random force.
Next consider the {\em propagation term}, which is the lhs of (\ref{a7}). 
In the limit of small $\delta t$ 
it simplifies to the usual streaming term $\partial_t f +
\mbox{\boldmath $v$} \cdot \mbox{\boldmath $\nabla$} f$, and the FPB equation 
becomes
\begin{equation} \label{a10}
\partial_t f(x) + \mbox{\boldmath $v$} \cdot \mbox{\boldmath $\nabla$} f(x) 
 = I(f) .
\end{equation}
Here $\mbox{\boldmath $\mbox{\boldmath $\nabla$}$} = \partial / 
\partial\mbox{\boldmath $r$}$
is a spatial gradient and the collision term becomes:
\begin{equation} \label{a11}
I(f) = \mbox{\boldmath $\partial$} \cdot \int dx^\prime
\mbox{\boldmath $\gamma$}(x,x^\prime ) f(x^\prime ) f(x) 
 +  \frac{1}{2} \mbox{\boldmath $\partial$}\mbox{\boldmath $\partial$} \; : 
\; 
\int dx^\prime \mbox{\boldmath $\sigma$} (x,x^\prime ) \mbox{\boldmath $\sigma$} 
(x,x^\prime) f(x^\prime) f(x).
\end{equation}

We will concentrate on the case of continuous time.
The most fundamental properties of the FPB equation are: it conserves
particle number $N= \int dx f(x,t)$ and total momentum $P = \int dx$
$\mbox{\boldmath $v$} f(x,t)$, and obeys an $H$-theorem, 
$\partial_t {\cal F} \leq 0$, for the total free energy:
${\cal F}(f) = \int dx \{ \frac{1}{2} mv^2 + \theta_0 \ln f(x) \} f(x) ,$
provided the {\em detailed balance} conditions \cite{E+W},
$\theta_0  = m\sigma^2 /2\gamma$ and 
$w_D (r)  =  w^{2}_{R} (r) \equiv w(r)$, are satisfied.
The constant $\theta_0$ depends only on the model parameters. In absence of conservative forces, the $H$-theorem implies the
existence of  a unique equilibrium solution, $f_0 (x) = n_0 \varphi_0(v)$
where $\varphi_0(v) = (m/2\pi\theta_0)^{d/2} \exp [- \frac{1}{2} 
mv^2/\theta_0]$
is a Maxwellian, and $n_0 = N/V$ is the density. 
If the detailed balance conditions are violated, the proof of the $H$-theorem 
breaks down.
The modified algorithm \cite{E+W} satisfies the detailed balance
conditions in the limit $dt \rightarrow 0$, but the original one \cite{H+K} does 
not.

Our goal here is to derive the macroscopic evolution equations, describing
the fluid dynamics on large spatial and temporal scales, i.e. the Navier-Stokes
equation, and to obtain explicit expressions for the thermodynamic and
transport properties in terms of density $n$, temperature $\theta_0$,
friction $\gamma$ and range $R_0$.

As the interactions in the DPD fluid do not conserve energy, only mass
density, $\rho (\mbox{\boldmath $r$},t) = mn(\mbox{\boldmath $r$},t)
=\int d \mbox{\boldmath $v$} f(\mbox{\boldmath $r$},\mbox{\boldmath $v$},t)$ 
, and 
momentum density $\rho (\mbox{\boldmath $r$},t) 
\mbox{\boldmath $u$}(\mbox{\boldmath $r$},t) = 
\int d \mbox{\boldmath $v$} f(\mbox{\boldmath $r$},
\mbox{\boldmath $v$},t) \mbox{\boldmath $v$},$ satisfy local
conservation laws,
where $\mbox{\boldmath $u$} (\mbox{\boldmath $r$},t)$ is the local 
flow velocity of the fluid. The approach to equilibrium proceeds in two
stages: first, a rapid relaxation ({\em kinetic stage}) to local equilibrium
within a characteristic kinetic relaxation time $t_0$. 
The energy density $e(\mbox{\boldmath 
$r$},t)$
decays within the same short kinetic stage to $\frac{1}{2} d\theta_0 
n(\mbox{\boldmath $r$},t)$,
where $\theta_0$ is the global equilibrium temperature $\theta_0 = k_B T_0$.
In the subsequent {\em hydrodynamic stage} the
time evolution of $f(x,t)$ in the DPD fluid is only determined by the slow fields 
$n(\mbox{\boldmath $r$},t)$
and $\mbox{\boldmath $u$}(\mbox{\boldmath $r$},t)$, i.e.
$f(\mbox{\boldmath $v$} | n (\mbox{\boldmath $r$},t), \mbox{\boldmath 
$u$}(\mbox{\boldmath $r$},t))$,
but the temperature is
thermostated at the global equilibrium value $\theta_0$. All processes
proceed isothermally. As a  DPD fluid is not able to sustain a temperature 
gradient
on hydrodynamic time scales, there is no heat current proportional to a
temperature gradient and there is no heat conductivity. 

Using the Chapman-Enskog method \cite{Chap},
$f$ can be determined perturbatively, $f = f_0 + \mu f_1 + \ldots $,
as an expansion in powers of a small parameter, $\mu \sim \ell_0 \nabla$, 
which
measures the variation of the macroscopic parameters over the 
characteristic kinetic length scale, $\ell_0 = \overline{v}t_0 \propto
\overline{v}/\gamma n$ with $\overline{v} = \sqrt{\theta_0/m}$ the mean
velocity.
The first term $f_0$ is the local equilibrium distribution, 
$f_0 = n(\mbox{\boldmath $r$},t) \left( m/(2 \pi \theta_0) \right)^{d/2} \exp 
\left[ - m
(\mbox{\boldmath $v$}-\mbox{\boldmath $u$}(\mbox{\boldmath $r$},t))^2/
(2\theta_0) \right].$
The next term $f_1$, linear in the gradients, is obtained as the solution of
$\partial_t f_0 + \mbox{\boldmath $v$} \cdot \mbox{\boldmath $\nabla$}
f_0 = (dI_0 /df)_{f_{0}} f_1$, where the time derivatives $\partial_t n$ and
$\partial_t \mbox{\boldmath $u$}$ are eliminated using the lowest order
hydrodynamic equations (Euler equations). So, we first need to consider the
local conservation laws.

Integration of (\ref{a10}) over $\mbox{\boldmath $v$}$ yields at 
once the continuity equation,
$\partial_t n = - \mbox{\boldmath $\mbox{\boldmath $\nabla$}$}\cdot n 
\mbox{\boldmath $u$}$.
Similarly, we derive the conservation law for the momentum density 
by multiplying
(\ref{a10}) by $\mbox{\boldmath $v$}$ and by integrating over
$\mbox{\boldmath $v$}$. The result after a partial $\mbox{\boldmath $v$}$-
integration is
\begin{eqnarray} \label{a16}
\partial_t \rho\mbox{\boldmath $u$} &=& - \mbox{\boldmath $\nabla$} 
\cdot
\int d\mbox{\boldmath $v$} \mbox{\boldmath $v$}\mbox{\boldmath $v$} f(x)
- m \int d \mbox{\boldmath $v$} dx^\prime \mbox{\boldmath $\gamma$} 
(x,x^\prime ) f(x^\prime ) f(x) \nonumber \\
& \equiv & - \mbox{\boldmath $\nabla$} \cdot (\rho \mbox{\boldmath $u$}
\mbox{\boldmath $u$} + \mbox{\boldmath $\Pi$}_{\rm K} + \mbox{\boldmath 
$\Pi$}_{\rm D}).
\end{eqnarray}
The kinetic part of the pressure tensor is the momentum flux in the local
rest frame of the fluid
$\mbox{\boldmath $\Pi$}_{\rm K} = \int d\mbox{\boldmath $v$} m \mbox{\boldmath 
$v$}\mbox{\boldmath $v$} 
f -
\rho \mbox{\boldmath $u$}\mbox{\boldmath $u$} = \int d\mbox{\boldmath $v$} m
\mbox{\boldmath $V$}\mbox{\boldmath $V$}f ,$
where $\mbox{\boldmath $V$} = \mbox{\boldmath $v$} - \mbox{\boldmath $u$} 
(\mbox{\boldmath $r$},t)$
is the {\em peculiar} velocity. Substitution of $f = f_0 + \mu f_1$ gives
to lowest order $\mbox{\boldmath $\Pi$}_{\rm K} \simeq n\theta_0 {\identity} + 
{\cal O}(\mu )$, where $n\theta_0$ is
the local equilibrium pressure.
The next term on the rhs of (\ref{a16}) defines the {\em dissipative} part
$\mbox{\boldmath $\Pi$}_D$ of the momentum flux. It reduces to 
\begin{equation}\label{a18}
\mbox{\boldmath $\nabla$} \cdot \mbox{\boldmath $\mbox{\boldmath $\Pi$}$}_{\rm D} 
= m\gamma \int 
d\mbox{\boldmath $R$} w(R) \widehat{{\bf R}}\widehat{{\bf R}} \cdot 
(\mbox{\boldmath $u$}(\mbox{\boldmath $r$}) -\mbox{\boldmath $u$}(\mbox{\boldmath 
$r$}^\prime )) n
(\mbox{\boldmath $r$}) n(\mbox{\boldmath $r$}^\prime ) ,
\end{equation}
where $\mbox{\boldmath $r$}^\prime = \mbox{\boldmath $r$} -
\mbox{\boldmath $R$}$. It is a typical {\em collisional transfer}
contribution, resulting from the nonlocality of the collision operator
\cite{Chap}. Expansion of $\mbox{\boldmath $u$}(\mbox{\boldmath $r$}^\prime )$ and 
$n(\mbox{\boldmath $r$}^\prime )$
around $\mbox{\boldmath $r$}$ shows that the rhs of (\ref{a18}) is 
${\cal O}(\mu^2 )$.  It contributes to the viscosities but not to the
Euler equations.  After some algebra we obtain
${\sf \Pi}_{\rm D} = -2\eta_{\rm D} {\sf D} - \zeta_{\rm D} \nabla
\cdot \mbox{\boldmath $u$} \identity$, where the dissipative parts of
the shear and bulk viscosity are identified as:
\begin{equation} \label{a24}
\eta_{\rm D} = m n \omega_0 \langle R^2\rangle_w/2(d+2) \qquad \mbox{ ; } \qquad
\zeta_{\rm D} =m n \omega_0 \langle R^2 \rangle_w/2d.
\end{equation}
where we define the kinetic relaxation rate
$\omega_0 \equiv 1/t_0 = (n\gamma/d) [w]$ with
$[w] = \int d{\bf R} w(R)$ the effective
volume of the action sphere, and where
$<R^2>_w = [R^2 w]/[w^2]$ is of order $R_0^2$.

Following the standard Chapman-Enskog scheme, the solution $f_1$ is found
to be proportional to $\mbox{\boldmath $\nabla$} \mbox{\boldmath $u$}$. The
corresponding part of the kinetic pressure tensor then has the form
\begin{equation} \label{a19}
\mbox{\boldmath $\Pi$}_{{\rm K},1} = \int d\mbox{\boldmath $v$}m \mbox{\boldmath 
$V$}
\mbox{\boldmath $V$} f_1 = - 2\eta_{\rm K}  {\sf D} - \zeta_{\rm K} 
\mbox{\boldmath $\nabla$} 
\cdot
\mbox{\boldmath $u$}  {\identity} ,
\end{equation}
where the rate of shear tensor ${\sf D}_{\alpha \beta}$ is defined as
the traceless symmetric part of $\nabla_{\alpha} \mbox{\boldmath $u$}_{\beta}$.
For the kinetic part of the viscosities we obtain the explicit results
\begin{equation} \label{a21}
\eta_{\rm K} = n\theta_0/2\omega_0 \qquad \mbox{ ; } \qquad \zeta_{\rm K} = 
n\theta_0/d\omega_0 ,
\end{equation}
which are new results.  The final transport coefficients are then
$\eta = \eta_{\rm D} + \eta_{\rm K}$ and $\zeta = \zeta_{\rm D}
+ \zeta_{\rm K}$.  For further details we refer to \cite{MBE}.

The contribution $\eta_D$ is
{\em identical} to the estimate for the total shear viscosity 
of the DPD fluid, calculated in \cite{H+K} on the basis of the continuum
approximation to the equations of motion for the DPD particles.
Hoogerbrugge and Koelman have also shown that the viscosity found in numerical 
simulations does indeed approach $\eta_D$ for large values of $n\gamma$. 
Here we have confirmed and rederived the formula for the viscosity of \cite{H+K} 
within a kinetic theory context. 

We conclude this paragraph by listing the complete results for the viscosities of
the DPD fluid in the continuous case $(\delta t \rightarrow 0)$,
\begin{equation} \label{a25}
\eta = \frac{1}{2} n\theta_0 \left\{ \frac{\omega_0
t^{2}_{w}}{(d+2)} + \frac{1}{\omega_0} \right\} \qquad \mbox{ ; } \qquad
\zeta  = \frac{1}{d} n\theta_0 \left\{ \frac{\omega_0
t^{2}_{w}}{2} + \frac{1}{\omega_0} \right\}.
\end{equation}
The traversal time of an action sphere $t_w$ is defined through
$t^{2}_{w} = m \langle R^{2}\rangle_w / \theta_0 = \langle R^2\rangle_w/
\overline{v}^2$. The expressions for the transport coefficients (\ref{a25}) 
involve the two intrinsic time scales of the DPD fluid: the characteristic 
kinetic
time $t_0 = 1/\omega_0$, and the traversal time $t_w$ which is of 
order
$R_0/\overline{v}$. In the parameter range $t_w > t_0$, 
the dissipative 
viscosities $\eta_{\rm D}$ and $\zeta_{\rm D}$ dominate and the estimate of 
\cite{H+K}
is a reasonable one. In the range $t_w < t_0$ the kinetic viscosities
$\eta_{\rm K}$ and $\zeta_{\rm K}$ dominate. 
In a similar way the coefficient of self-diffusion $D$ is obtained as  
$D=\theta_0/\omega_0 m$.

Results of numerical simulations for the kinematic viscosity
$\zeta/(nm)$ show \cite{MBE} that the theoretical results correctly
describe the large and small $\omega_0 \sim n\gamma$ dependence, but a
background contribution, roughly independent of $\omega_0$ seems to be
lacking in the theoretical predictions.

\noindent
There may be several reasons for this discrepancy, which are currently
under investigation:\\ 
(i) Effects of the discreteness of $\delta t$, which alter the
equilibrium temperature \cite{M+Y}.\\
(ii) Poor convergence of the Chapman-Enskog expansion, resulting from a
poor separation of the ``fast'' kinetic and ``slow'' hydrodynamic
timescales.\\
(iii) The system size L, measured in units of the interaction range
$R_0$, as used in the simulations of \cite{MBE}, may be too small and
one is observing finite size effects related to wavenumber dependent
transport coefficients ( generalized hydrodynamics).\\
(iv) Breakdown of the stosszahlansatz, caused by the {\em small} net
momentum transfer in DPD collisions.  Consequently, dynamic
correlations, built up by sequences of correlated binary (``ring'')
collisions, may not be negligible.

The main results of the present paper are the explicit predictions for the
viscosities and self-diffusion coefficient of the DPD fluid in terms
of the model parameters: density $n$,
friction $\gamma$, noise strength $\sigma$ or equivalently temperature
$\theta_0 = m\sigma^2 /2\gamma$ and range $R_0$. The
expressions for $\eta$ and $\zeta$ essentially depend on the two intrinsic time
scales of DPD: the characteristic kinetic relaxation time $t_0$ and the 
traversal time $t_w$ of the action sphere.

The method used in eqs. (\ref{a5}-\ref{a7}) for deriving  equations of 
motion 
for the reduced distribtion functions, can be directly applied to the $N$-particle 
distribution function, $P(y_1, y_2,\ldots , y_N,t) = \langle \mbox{\boldmath 
$\Pi$}^{N}_{i=1} \delta (y_i - x_i (t)) \rangle$, and yields the discrete time 
analog of the $N$-particle Fokker-Planck equation for continuous time \cite{E+W}. 
Our equation for finite step size
reduces to the Fokker-Planck equation of \cite{E+W} in the limit $\delta t 
\rightarrow 0$.

\section*{Acknowledgements}
C. Marsh acknowledges support from EPSRC (UK), Unilever Research (UK)
and ERASMUS (EU) as well as the hospitality of Universiteit Utrecht.
G. Backx is a post-doctoral researcher for the foundation ``Fundamenteel
Onderzoek der Materie'' (FOM), which is financially supported by the
Dutch National Science Foundation (NWO).


\begin{thebibliography}{99}

\bibitem{H+K} 
P.J. Hoogerbrugge and J.M.V.A. Koelman, Europhys.Lett. {\bf 19}, 155 (1992).

\bibitem{E+W}
P. Espa\~nol and P. Warren, Europhys.Lett. {\bf 30}, 191 (1995).

\bibitem{Colloid}
J.M.V.A. Koelman and P.J. Hoogerbrugge, Europhys. Lett. {\bf 21}, 363 (1993);
E.S. Boek, P.V. Coveney and H.N.W. Lekkerkerker,
J.Phys.Cond.Matt. {\bf 8}, 9509 (1996); A.G. \newline Schlijper, P.J. Hoogerbrugge 
and C.W. 
Manke, J. Rheol. {\bf39}, 567 (1995); P.V. Coveney and K.E. Novik, Phys.Rev.E {\bf 
54}, 5134 (1996).

\bibitem{M+Y}
C. Marsh and J. Yeomans, Europhys. Lett. {\bf 37} (8), 511 (1997)

\bibitem{Chap}
S. Chapman and T.G. Cowling.  The Mathematical Theory of Non-Uniform
Gases (Cambridge University Press, 3rd edition 1970).

\bibitem{MBE} 
C. Marsh, G. Backx, and M.H. Ernst, WWW cond-mat/9702036.

\end{thebibliography}
 \end{document}